\documentclass[letterpaper,11pt]{article}
\usepackage{amsmath,amssymb,amsthm,tikz}    
\usepackage{bridges}			
\usepackage{graphicx}			
\usepackage[colorlinks=true, urlcolor=blue, citecolor=black, linkcolor=black]{hyperref}  
\usepackage{subcaption}			

\title{Automating Crochet Patterns for Surfaces of Revolution}
\author{Megan Martinez\textsuperscript{1} and Amanda Taylor Lipnicki\textsuperscript{2}
\vspace{10pt}\\
\textsuperscript{1}Mathematics Department, Ithaca College; mmartinez@ithaca.edu\\
\textsuperscript{2}Math and CS Division, Alfred University; tayloral@alfred.edu} 

\date{}					

\begin{document}

\maketitle

\thispagestyle{empty}

\begin{abstract}

A surface of revolution is created by taking a curve in the $xy$-plane and rotating it about some axis. We develop a program which automatically generates crochet patterns for surfaces by revolution when they are obtained by rotating about the $x$-axis. In order to accomplish this, we invoke the arclength integral to determine where to take measurements for each row. In addition, a distance measure is created to optimally space increases and decreases. The result is a program that will take a function, $x$-bounds, crochet gauge, and a scale in order to produce a polished crochet pattern.
\end{abstract}



\section*{Introduction}

Physical models of mathematical objects allow the theoretical to become tangible. This project began as a way to help students visualize and engage with the quadric surfaces by creating crochet models. As we tinkered with patterns and calculations, we realized our methods were developing an algorithm to produce crochet patterns for surfaces by revolution. We fully developed our methods into a CoCalc worksheet that automatically produces crochet patterns for surfaces of revolution. In this paper, we explain the methods we used to develop this program.

Surfaces by revolution in fiber arts is not a new concept. The question of creating knitting patterns to construct surfaces of revolution was introduced by Amy F. Szczepa\'{n}ski \cite{Szczepanski}, who developed a method for giving the number of stitches that should appear in each row of a pattern. Our work is different from Szczepa\'{n}ski's in a few ways: we use different methods for determining the number of stitches in a row and have taken the process further to produce a polished crochet pattern. Our output should look like something that you would read on Ravelry.com (a popular repository of knitting and crochet patterns). 

We endeavored to produce crochet instructions that would create a physical model as true to the theoretical one as possible. This includes the following considerations:
\begin{itemize}
    \item Planning where to measure circumference of a surface to emphasize local minima and maxima
    \item Spacing increases and decreases to minimize distortion, while keeping the pattern easy to crochet
    \item Identifying what behaviors are problematic for our program
\end{itemize}

Note that while our pattern specifically creates a crochet pattern, it would be relatively straightforward to tinker with the output to read like a knitting pattern. Practically, there is not enough different between the mechanics of the two crafts to greatly affect the work we have done.

\subsection*{Accessing the Code}

Our code was written in a Sage worksheet, so can be freely accessed using CoCalc. We have published the code in the github repository \url{https://github.com/meganmartinez/math_crochet}. You can easily bring this code into CoCalc here: \url{https://cocalc.com/github/meganmartinez/math_crochet}. The ``README.md'' file has basic instructions for operating the code. If you click on ``Crocheting Surfaces of Revolution (v13).sagews'' you will see the code, but you must click ``edit'' to be able to evaluate the code and work on your own patterns. You can do this without a CoCalc account, though creating one is free.

\subsection*{Inputs \& Crochet Terminology}

Our program takes the following inputs:
\begin{itemize}
    \item $f(x)$ is the function that will be rotated about the $x$-axis to create the surface. The function $f(x)$ should be positive on $(a,b)$ and $f'(x)$ should be defined on $[a,b]$ in order for our code to function.
    \item $a$ is the $x$-value determining the start of the surface
    \item $b$ is the $x$-value determining the end of the surface.
    \item $S$ is the stitch gauge; that is, the number of stitches that fit in 4"
    \item $R$ is the row gauge; that is, the number of rows that fit in 4"
    \item $scale$ is the measure of one unit in inches (so allows one to decide how large the model will be)
\end{itemize}

Note we have worked gauge, the size of an individual's crochet stitch, into this program. In order to achieve the most accurately proportioned surface, it is recommended that the user makes a gauge swatch using their yarn and crochet hook of choice. We have worked to minimize the techniques a crocheter needs to be familiar with to follow our patterns. In service to this, all of our shapes are crocheted in-the-round using the spiral method. However, the experienced crocheter is encouraged to alter this using a joining method if desired. We use the following crochet techniques and symbols:

\begin{itemize}
    \item Chain
    \item Sc: Single crochet. Sc5 would mean ``single crochet 5 stitches''
    \item Inc: Make two crochet stitches in one stitch; turns one stitch into two.
    \item Dec: Invisible decrease; turns two stitches into one.
\end{itemize}

\section*{Identifying Row Landmarks}

The first order of business for our program is to identify where the circumference of the surface will be measured. To this end, we identify $x$-values (which we call landmarks) where we will measure the circumference of our surface. These will correspond to each row of the crochet pattern.

In her work, Szczepa\'{n}ski does this by approximating the curve defining the edges of a shape by using line segments of a fixed length, $h$. If one of your endpoints of a line segment is $(x_i,y_i)$, then your other endpoint would be where the curve intersects $(x-x_i)^2+(y-y_i)^2=h^2$ (something that can be solved or approximated by computer).

We take a slightly different approach by using arclength calculations as the basis for where we place the rows, as opposed to using secants in the method of Szczepa\'{n}ski. This allows us to more easily use our program to set rows at important landmarks of a function, as we will see in the next section. 


Consider the following values: $f(x)=x^3+2x^2-2x+4$, $a=-3$. $b=1$, $S=22$, $R=25$, $scale=0.18$.

Since our goal is to output a crochet pattern, it is helpful to measure distances using the units of ``rows'' or ``stitches.'' For instance, the arclength of our curve in rows is \[L=\frac{scale \cdot R}{4}\int_{a}^b \sqrt{1+(f'(x))^2}\,dx=\frac{(0.15)(25)}{4}\int_{-3}^1 \sqrt{1+(3x^2+4x-2)^2}\,dx\approx 16.162 \text{ rows}.\] Recognize that $scale$ is in inches/unit, $\frac{R}{4}$ is in rows/inch and our integral is in units.

\begin{figure}[h!tbp]
\centering
\begin{minipage}[b]{0.3\textwidth} 
	\includegraphics[width=\textwidth]{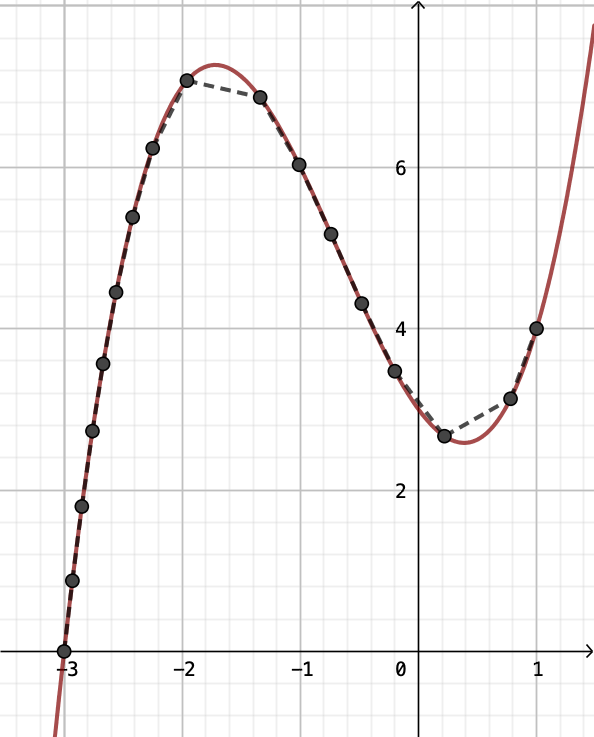}
        	\subcaption{} 
         \label{Graph1}
\end{minipage}
\quad\quad
\begin{minipage}[b]{0.3\textwidth} 
	\includegraphics[width=\textwidth]{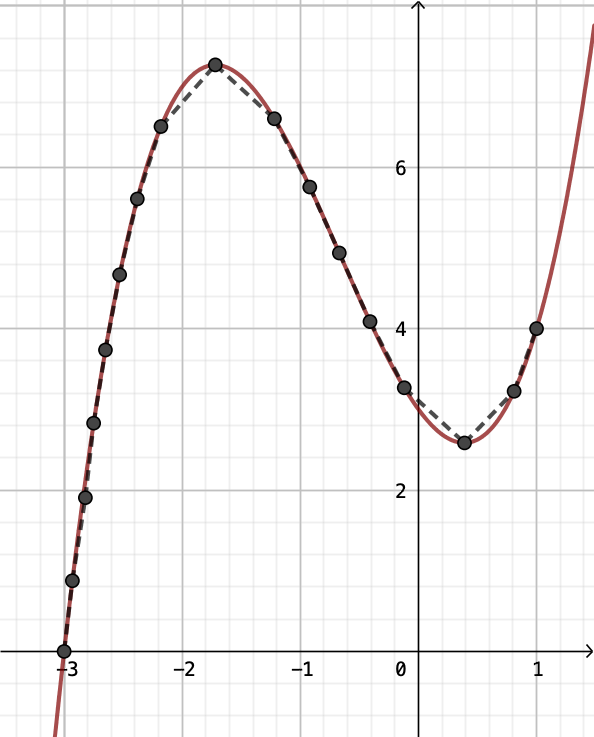}
        	\subcaption{} 
         \label{Graph2}
\end{minipage}
\caption{The function $x^3+2x^2-2x+4$ plotted with (a) the 13 $x$-values where evenly-space rows land and (b) the more refined approach of determining these $x$-values while considering the location of local extrema.}
\end{figure}

We round this to 16 rows in our final pattern. Since we rounded, we can try and minimize the effect of that rounding by evenly placing those 16 rows along the curve. We place an initial cast-on/chain row at $x_0=-3$, then include 16 more rows of instruction. The next row, $x_1$ will be $\frac{16.162}{16}\approx 1.010$ distance down the curve. In general \[\frac{scale \cdot R}{4}\int_{x_0}^{x_{i}} \sqrt{1+(f'(x))^2}\,dx = i\cdot \dfrac{L}{[L]}.\] Note that we are using $[x]$ to indicate $x$ rounded to the nearest integer. In general, this isn't a particularly ``nice'' equation to solve. In order to make a program that will work for the most functions possible, we implemented a ``guess and check'' protocol to find each $x_i$ accurate to two decimal places. To mitigate the effects of rounding error, we always measure our arclength from $x_0 = a$.

So, in our current example, we have rows at the following $x$-landmarks: $x_0=a=-3, x_1=-2.93, x_2 = -2.85, x_3=-2.76,x_4=-2.67, x_5=-2.56, x_6=-2.42, x_7=-2.25, x_8=-1.96, x_9=-1.34, x_{10}=-1.01, x_{11}=-0.74, x_{12}=-0.48, x_{13}=-0.20, x_{14}=0.22, x_{15}=0.78, x_{16}=b=1.00$. Each of these values have been plotted in Figure~\ref{Graph1}. Notice that there are local extrema at $\dfrac{-2\pm\sqrt{10}}{3}$ that are between our rows. This means that when knit or crocheted, our model will have slightly less accentuated peaks and valleys. As we increase the value of $scale$ in our program, we will work more rows and this effect will be minimized. As we decrease $scale$, the problem will be worse.

\subsection*{Prioritizing Local Extrema}

Our method of using arclengths allows us to make slight adjustments to our $x$-values to accentuate the local extrema of a shape. We simply do the following:

\begin{enumerate}
    \item Create a list including the endpoints, $a$ and $b$, and all the $x$-values of local extrema for your given function between $a$ and $b$: $\{m_0=a,m_1,\ldots,m_n=b\}$. 
    \item Between each $m_i$ and $m_{i+1}$, find the $x$-values for where to place rows as in the previous section.
\end{enumerate}

\begin{table}
\centering
\begin{tabular}{|c|c|c|}
\hline
$i$ & $(cR/4)\int_{m_i}^{m_{i+1}}\sqrt{1+(3x^2+4x-2)^2}\,dx$ & $x$-Landmarks \\\hline
0 & 8.434 & $x_0=m_0=-3, x_1=-2.93, x_2=-2.84, x_3=-2.75, x_4=-2.65,$ \\
& & $ x_5=-2.53, x_6=-2.38, x_7=-2.18, x_8=m_1=(-2-\sqrt{10})/3$ \\\hline
1 & 5.923 & $x_0=m_1=(-2-\sqrt{10})/3, x_1=-1.22, x_2=-0.92, x_3=-0.67,$ \\
 & & $x_4=-0.41, x_5=-0.12, x_6=m_2=(-2+\sqrt{10})/3$\\\hline
2 & 1.805 & $x_0=m_2=(-2+\sqrt{10})/3, x_1=0.81, x_2=m_3=1$\\\hline
\end{tabular}
\caption{Calculations adjusting $x$-landmarks.}
\label{CalcTable}
\end{table}

Continuing with our example, we construct the list of extrema $\{m_0=-3, m_1=\frac{-2-\sqrt{10}}{3}, m_2=\frac{-2+\sqrt{10}}{3},\allowbreak m_3=1\}$. We then find $x$-landmarks between each $m_i$ value in Table~\ref{CalcTable}.
\def\arraystretch{1.75}

Taken together, we get a new list of $x$-landmarks:
$-3,-2.93,-2.84,-2.75,-2.65,-2.53,-2.38,-2.18,\allowbreak\dfrac{-2-\sqrt{10}}{3},\allowbreak-1.22,-0.92,-0.67,-0.41,-0.12,\dfrac{-2+\sqrt{10}}{3},0.81,1$. These are pictured in Figure~\ref{Graph2}.

\subsection*{Calculating Number of Stitches in Each Row}

Once the $x$-landmarks have been determined for the placement of each row, it is a simple matter to calculate how many stitches should be in each row. There are $s(i)=[2\pi \cdot scale \cdot\frac{S}{4}f(x_i)]$ stitches in the $i$-th row. In our running example, the 2nd row has $[2\pi(0.18)\cdot\frac{22}{4}f(-2.84)] = 18$ stitches. The number of stitches in each row of our example have been displayed in Table~\ref{stitches}.

\def\arraystretch{1}
\begin{table}
\centering

\begin{tabular}{|c|c||c|c||c|c|}
\hline
Row & \# Stitches & Row & \# Stitches & Row & \# Stitches \\\hline
0 & 6 & 6 & 41 & 12 & 32 \\\hline
1 & 12 & 7 & 47 & 13 & 27 \\\hline
2 & 18 & 8 & 51 & 14 & 22 \\\hline
3 & 24 & 9 & 47 & 15 & 26 \\\hline
4 & 29 & 10 & 42 & 16 & 31 \\\hline
5 & 35 & 11 & 37 &  &  \\\hline
\end{tabular}
\caption{The number of stitches in each row of our pattern.}
\label{stitches}
\end{table}

\section*{Writing the Pattern}

While computing the number of stitches in each row of our pattern is a big part of pattern construction, there are many additional considerations for how to output a polished set of instructions.

\subsection*{Increase and Decrease Placement}
The most important problem to solve is how to distribute increases and decreases in an effective, yet practical way. When increases or decreases are placed on top of each other in consecutive rows, it can affect the smoothness of the finished shape. However, if we have the program output instructions that are burdensome to follow, then it will be challenging for the user to follow the instructions. We will discuss each of these considerations and how we decided to balance the two.

Suppose we are considering Row 6 from above with $s(6)=41$ stitches. Row 5 has $s(5)=35$ stitches, so Row 6 will need to include 6 increase stitches. Following are some options for how those could be distributed. Note that we use the symbol * to block off a section of instructions that will be repeated a designated number of times. 

\begin{description}
\item[Option 1:] Sc35, *Inc* (6 times)
\item[Option 2:] *Sc4, Inc* (6 times), Sc5
\item[Option 3:] *Sc4, Inc* (2 times), Sc1, *Sc4, Inc* (2 times), Sc2, *Sc4, Inc* (2 times), Sc1
\end{description}

One could argue that the first set of instructions is the easiest, but by grouping all the increase stitches together, we are creating a skew that will pull our finished surface to one side, essentially bending the $x$-axis in our physical model. Options 2 and 3 do a much better job of spreading the increases or decreases along the row. Option 2 is computed with a simple division calculation (which we will formalize below)---we call this the \emph{remainder method}. For this option, we have spread out the increases a fair amount, but tacked on additional ``remainder'' stitches at the end of the row.

Option 3 works to additionally split up that clump of non-decrease stitches at the end of the row, by distributing those as well. It is easily argued that something like Option 3 would be the best choice theoretically; however, we must always keep in mind that a real person will be following these instructions, so we want to minimize how much counting and pattern-reading the crocheter has to do. In our program, we opt for the remainder method in Option (2) not because it is the most even distribution of the increases or decreases, but because it makes an attempt at even distribution while maintaining the sanity of the crocheter. Our decision means there is the possibility for a small amount of leaning in our shape (like in Option 1 to a much smaller degree). This effect will depend on how large the remainders are for a particular shape. From our experience, this effect is unnoticeable.

The other benefit to the remainder method is that we can easily use the remainders to shift the increases or decreases around the row and minimize the number of increase/decrease stitches that line up from row to row. For instance, instead of ``*Sc4, Inc* (6 times), Sc5,'' we could choose to make the row instructions ``Sc3, *Sc4, Inc* (6 times), Sc2'' or even ``Sc2, Inc, *Sc4, Inc* (5 times), Sc7''; notice in this last example, we pull from some of the single crochet stitches not in the remainder. To figure out the optimal configuration of the stitches, we create a distance measure.

For each row, we can create a set of ratios that indicate where the increases and decreases fall. For instance, ``*Sc4, Inc* (6 times), Sc5'' will have the associated set of ratios: $\{5/35, 10/35, 15/35, 20/35, 25/35, \allowbreak30/35\}$. Notice that the denominator in these ratios corresponds to the number of instructions in a row (as opposed to the number of stitches). For instance, ``*Sc4, Inc* (6 times), Sc5'' ends with 41 stitches, and is working 35 stitches from the previous row, while ``*Sc4, Dec* (6 times), Sc5'' ends with 35 stitches and is working 41 stitches from the previous row. For both, we would have the same set of ratios. Hence, for the denominator of the ratios in row $i$, we use $\min(s(i-1),s(i))$.

In general, we will write $D(i)$ to denote the positions of the increases/decreases in the $i$-th row. If $D$ gives some set of positions for the increases or decreases in the $i$-th row, we calculate the corresponding set of ratios with $r_i(D)=\{x/\min(s(i-1),s(i)) \mid x \in D\}$. Using these ratios, we create two types of distance measures between two consecutive rows. If $D_1$ (resp.\ $D_2$) is some set of increase/decrease positions in row $i-1$ (resp.\ $i$), then \[d_1(r_{i-1}(D_1),r_i(D_2))= \min\{\min(|y-x|, 1-|y-x|) \mid x \in r_{i-1}(D_1), y \in r_{i}(D_2)\}\] \[d_2(r_{i-1}(D_1),r_i(D_2))= \frac{1}{|D_1||D_2|}\sum_{x \in r_{i-1}(D_1), y \in r_{i}(D_2)}\min(|y-x|, 1-|y-x|)\]

The $d_1$ measure tells you the smallest pairwise distance that occurs between the ratios of the two rows, while the $d_2$ measure calculates the average pairwise distance between these ratios. Note that we write $\min(|y-x|, 1-|y-x|)$ to find the distance between two particular ratios; we need to account for the fact that one could be almost at the start of a row and while the other could be near the end, which actually makes them physically near each other (even if numerically they aren't).


With these distances defined, our pattern creation program follows the following algorithm to find the optimal placement of increases and decreases. Define integers $q$ and $r$ such that $0 \leq r < |s(i)-s(i-1)|$ (our remainder) and $\min(s(i-1),s(i))=q \cdot |s(i)-s(i-1)|+r$. Note that $|s(i)-s(i-1)|$ gives the number of increases or decreases needed in a row, while $\min(s(i-1),s(i))$ gives the number of instructions needed. We then construct $D(i)$, giving the positions of the increases or decreases in the $i$th row, as follows:

\begin{enumerate}
\item Set $desired\_positions=\{q\cdot j + 1 \mid 0 \leq j < |s(i)-s(i-1)| \}$ (note that this will group the $r$ remainder stitches at the end of the row). Set $prevdistance=0$ and $prevmean=0$.
\item For $k$ where $1 \leq k \leq q+r$:
\begin{enumerate}
\item Set $D'(i)=\{q\cdot j + k \mid 0 \leq j < |s(i)-s(i-1)| \}$, $newdistance = d_1(r_{i-1}(D(i-1)),r_{i}(D'(i)))$ and $newmean = d_2(r_{i-1}(D(i-1)),r_i(D'(i)))$
\item If $newdistance>prevdistance$ or ($newdistance = prevdistance$ and $newmean>prevmean$), set $prevdistance = newdistance$, $prevmean = newmean$, and $desired\_positions = D'(i)$
\end{enumerate}
\item Return $desired\_positions$
\end{enumerate}

Our output from this process will be the positions of increases and decreases that maximizes the $d_1$ distance. In the case of a tie, we choose the option which has the largest $d_2$ distance. This is especially important in those situations where $d_1$ is always 0 (meaning it is impossible to avoid an increase or decrease aligning with one in the previous row); in these situations, we just try to get everything else as far apart as possible.

For instance, in our running example, the Row 5 instructions are ``Sc6, Inc, *Sc3, Inc* (5 times), Sc2.'' and $r_5(D(5))=\{7/29, 11/29, 15/29, 19/29, 23/29, 27/29\}$. Our program will then run through all the lines in Table \ref{Inc_Calc} to find the optimal placement of increases in Row 6. The largest $d_1$ distance is produced with the last set of instructions, so we choose those.


\begin{table}
\centering
\resizebox{1\textwidth}{!}{
\begin{tabular}{|c|c|c|c|}\hline
Instructions & $r_6(D'(6))$ & $d_1(r_5(D(5)),r_6(D'(6)))$ & $d_2(r_5(D(5)),r_6(D'(6)))$ \\\hline
Sc5, *Sc4, Inc* (6 times) &  $\{10/35, 15/35, 20/35, 25/35, 30/35, 35/35\}$ & 0.04433& 0.05665\\\hline
Sc4, *Sc4, Inc* (6 times), Sc1 & $\{9/35, 14/35, 19/35, 24/35, 29/35, 34/35\}$ & 0.01576&0.02808 \\\hline
Sc3, *Sc4, Inc* (6 times), Sc2 & $\{8/35, 13/35, 18/35, 23/35, 28/35, 33/35\}$ & 0.00197&0.00739 \\\hline
Sc2, *Sc4, Inc* (6 times), Sc3 & $\{7/35, 12/35, 17/35, 22/35, 27/35, 32/35\}$ & 0.01675&0.02906 \\\hline
Sc1, *Sc4, Inc* (6 times), Sc4 & $\{6/35, 11/35, 16/35, 21/35, 26/35, 31/35\}$ & 0.04532 & 0.05764 \\\hline
*Sc4, Inc* (6 times), Sc5 & $\{5/35, 10/35, 15/35, 20/35, 25/35, 30/35\}$ & 0.04433 & 0.06158 \\\hline
Sc3, Inc, *Sc4, Inc* (5 times), Sc5 & $\{4/35, 9/35, 14/35, 19/35, 24/35, 29/35\}$ & 0.01576 & 0.04253 \\\hline
Sc2, Inc, *Sc4, Inc* (5 times), Sc6 & $\{3/35, 8/35, 13/35, 18/35, 23/35, 28/35\}$ & 0.00197 & 0.03120 \\\hline
Sc1, Inc, *Sc4, Inc* (5 times), Sc7 & $\{2/35, 7/35, 12/35, 17/35, 22/35, 27/35\}$ & 0.02167 & 0.04729 \\\hline
Inc, *Sc4, Inc* (5 times), Sc8 & $\{1/35, 6/35, 11/35, 16/35, 21/35, 26/35\}$ & 0.05025 & 0.06634 \\\hline
\end{tabular}
}
\caption{An example of the distance calculations to determine the optimal placement of increases or decreases in a row.}
\label{Inc_Calc}
\end{table}

\subsection*{Roots at $a$ or $b$}

Naturally, we don't want to consider functions that have a root between $a$ and $b$, since this doesn't make for a good physical model. However, there is no reason that we cannot create models for functions that have roots \emph{at} $a$ or $b$. To this end, we have created instructions that will instruct the crocheter to either begin or end a closed shape when appropriate. If there is a root at both $a$ and $b$, we further include instructions for stuffing the shape with fiberfill---a mathematical plushie!

\section*{Output}

Given $f(x)=x^3+2x^2-2x+4$, $a=-3$. $b=1$, $S=22$, $R=25$, $c=0.18$, our program produces the following pattern.

\begin{verbatim}
Row 0: Chain 6. join work, and Sc6.
Row 1:  *Inc* (6 times). (12 stitches)
Row 2:  Inc, *Sc1, Inc* (5 times), Sc1. (18 stitches)
Row 3:  *Sc2, Inc* (6 times). (24 stitches)
Row 4:  Sc1, Inc, *Sc3, Inc* (4 times), Sc6. (29 stitches)
Row 5:  Sc6, Inc, *Sc3, Inc* (5 times), Sc2. (35 stitches)
Row 6:  Inc, *Sc4, Inc* (5 times), Sc9. (41 stitches)
Row 7:  Sc3, Inc, *Sc5, Inc* (5 times), Sc7. (47 stitches)
Row 8:  Sc12, Inc, *Sc10, Inc* (3 times), Sc1. (51 stitches)
Row 9:  Sc6, Dec, *Sc10, Dec* (3 times), Sc7. (47 stitches)
Row 10:  Sc4, Dec, *Sc7, Dec* (4 times), Sc5. (42 stitches)
Row 11:  Dec, *Sc6, Dec* (4 times), Sc8. (37 stitches)
Row 12:  Sc3, Dec, *Sc5, Dec* (4 times), Sc4. (32 stitches)
Row 13:  Dec, *Sc4, Dec* (4 times), Sc6. (27 stitches)
Row 14:  Sc2, Dec, *Sc3, Dec* (4 times), Sc3. (22 stitches)
Row 15:  Sc6, Inc, *Sc4, Inc* (3 times). (26 stitches)
Row 16:  Sc1, Inc, *Sc4, Inc* (4 times), Sc4. (31 stitches)
Tie off
\end{verbatim}

\begin{figure}[h!tbp]
\centering
\begin{minipage}[b]{0.17\textwidth} 
	\includegraphics[width=\textwidth]{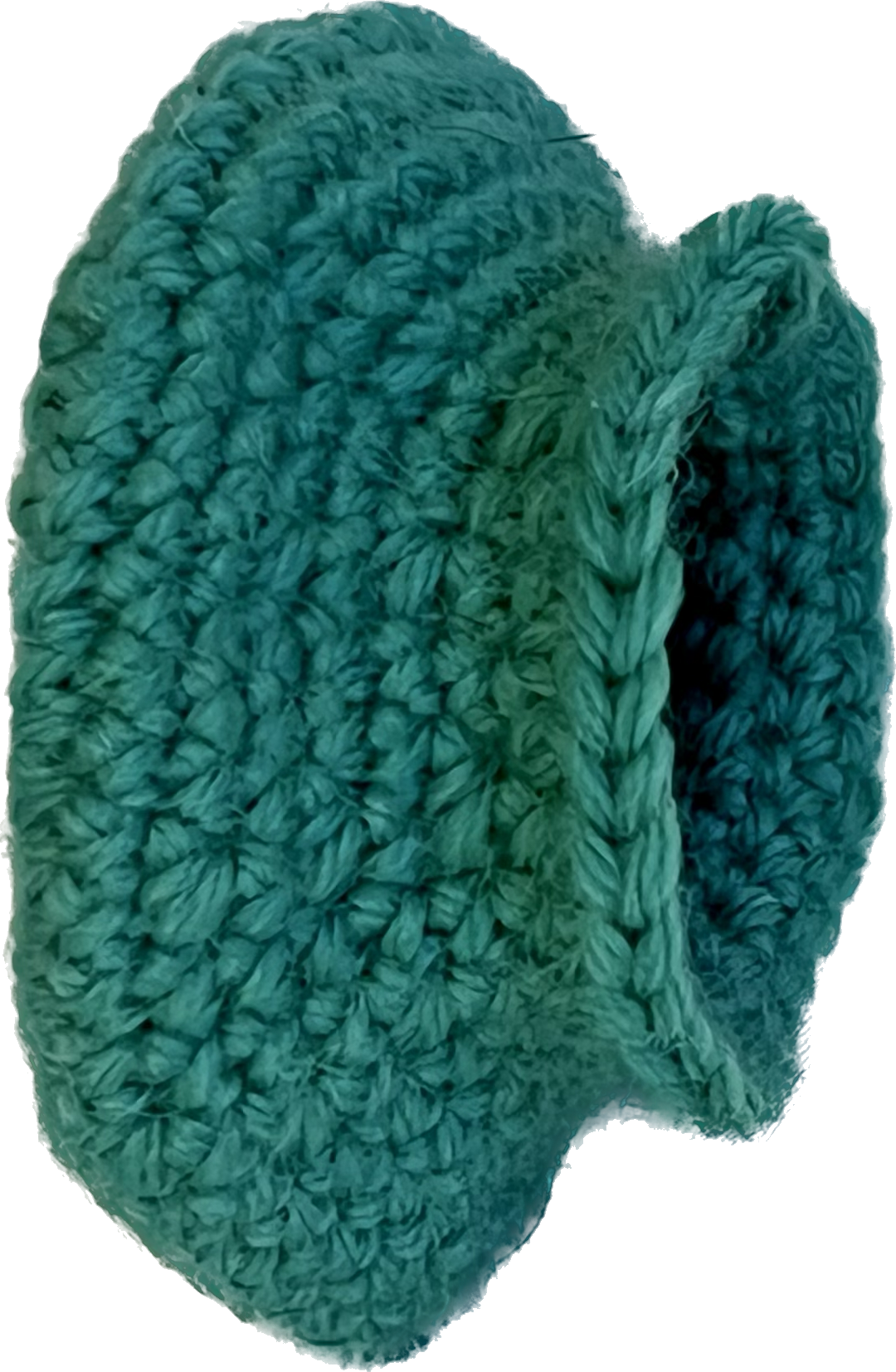}
        	\subcaption{} 
\end{minipage}
\quad	
\begin{minipage}[b]{0.13\textwidth} 
	\includegraphics[width=\textwidth]{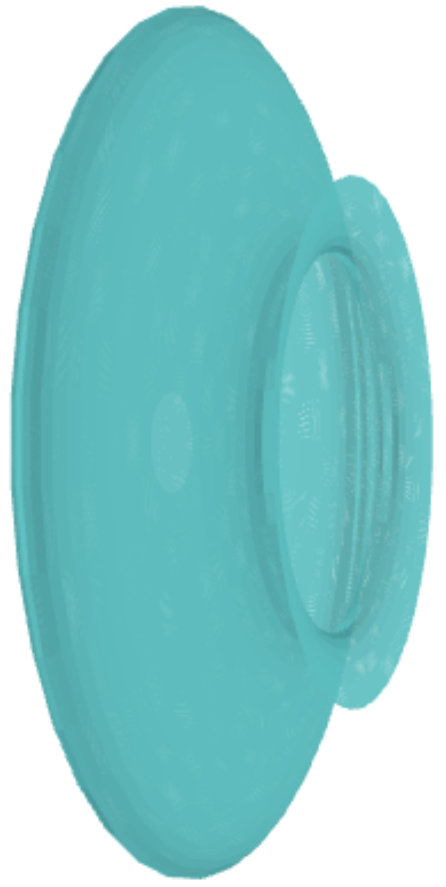}
        	\subcaption{} 
\end{minipage}
\quad
\begin{minipage}[b]{0.17\textwidth} 
	\includegraphics[width=\textwidth]{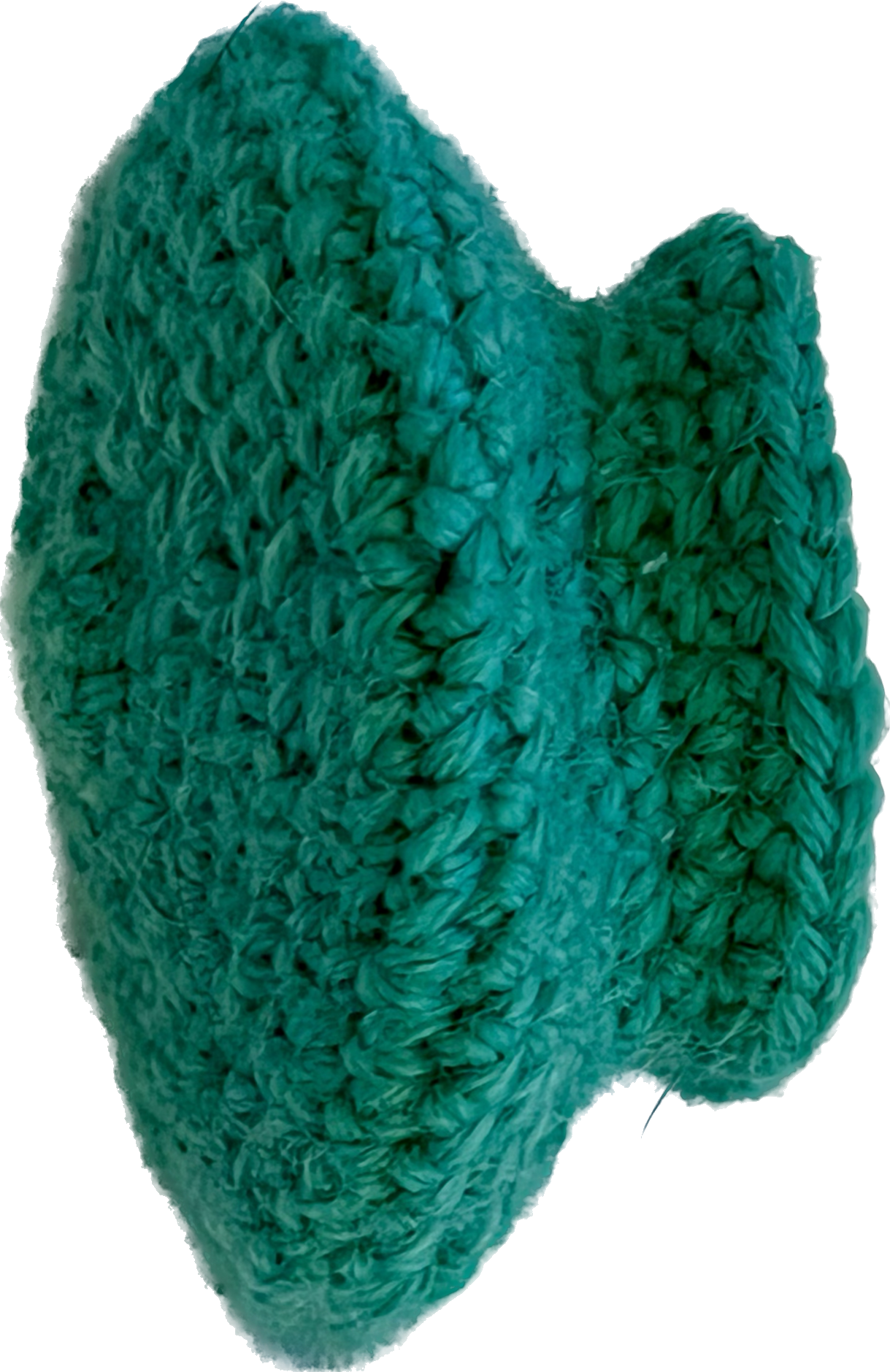}
        	\subcaption{} 
\end{minipage}
\quad	
\begin{minipage}[b]{0.11\textwidth} 
	\includegraphics[width=\textwidth]{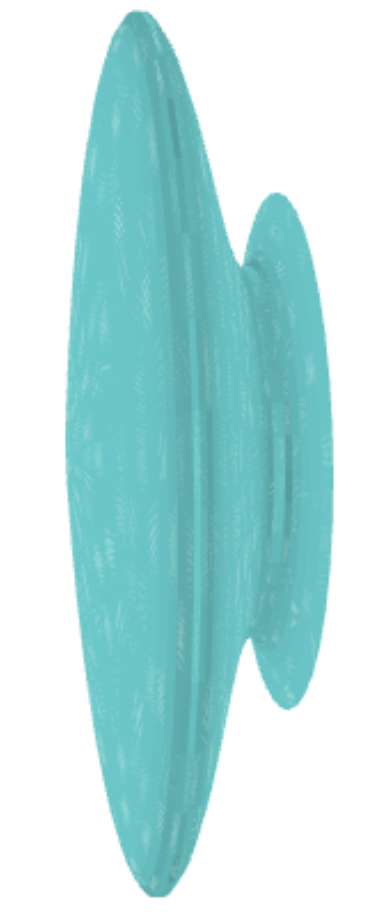}
        	\subcaption{} 
\end{minipage}

\caption{Figures (a) and (c) show different views of our crocheted model, using the instructions produced by our program, while (b) and (d) show our surface of revolution as pictured by CoCalc.}
\label{fig:crochet}
\end{figure}

The result of following these instructions is in Figure~\ref{fig:crochet}. Note that we have crocheted a relatively small shape here; the larger the shape, the closer it will look to the idealized picture.

\section*{Program Limitations}
The various decisions we made in the construction of our pattern program leads to limitations in what can be accomplished. In this section, we discuss a few of those limitations. Ultimately, it is fairly easy to choose a set of inputs that will not lead to a good pattern. We have tried to mitigate this issue by encoding some feedback from the program for when the user should reconsider their input.

It should be clear that crocheting impossibly complicated functions (such as $\sin(1/x)$ anywhere near $x=0$) will not work. The user should use some amount of judgement in deciding what model is reasonable to make. However, besides this ``common sense'' advice, there are some situations that will cause errors in the program.


\subsection*{Steep Increases or Decreases}
On occasion, there will be a change in stitch count from row to row that either more than doubles or more the halves the number of stitches. In this case, using Inc or Dec is not enough to accomplish this move. Our program will recognize this problem and make a note of it at the top of the pattern if it has occurred. Most of the time, this occurs when the function gets close to the $x$-axis. There are a couple ways to change your inputs to avoid this issue:

\begin{itemize}
\item Add a positive constant to your chosen function. There is always a constant that will solve the problem, since adding such a constant will increase the number of stitches to work with, but will not change the difference in stitches between rows.
\item Sometimes a change to $scale$ or multiplying your function by a constant can help, but this will not always work to solve the problem when your function gets very close to zero.
\end{itemize}

\subsection*{Approximation}

To create a program that works for more functions, we used ``guess and check'' to solve some equations to an accuracy of 0.01. In many cases, this works quite well. In others, it creates problems. In particular, using functions with features that are closer together that 0.01 will not be noticed by our program. We do not recommend that such functions are used.

When functions have a very rapid increase, there may be two $x$-landmarks that are exactly the same to two decimal places, or end up looking unevenly spaced because of the rounding. Since we always measure our arclength from the beginning of the shape, this kind of problem gets corrected and we rely on the forgiving nature of crochet to make a quality shape.

\section*{What More}

There are so many mathematical objects that can be worked into crochet---what other classes of mathematical objects might be amenable to computer generation? Our program is capable of producing a pattern for every quadric surface, besides the hyperbolic paraboloid. While we have devised a pattern for this surface \cite{Taylor}, we ask: is there a way to automate a pattern creator for surfaces that don't have circular cross-sections?

At a number of crossroads in the coding of this program, we made subjective decisions on how to proceed. Deciding where to place row landmarks, how to distribute increases and decreases, and even how to compute essential numbers, could all be done a different way, to different effect. Our aim has been to make a program that is broadly applicable, reasonably fast, and produces a pattern that would be enjoyable to crochet. We invite interested parties to make changes as they see fit to highlight the considerations they most care about. And of course, we invite crocheters to share their mathematical creations that arise from this program!

    
{\setlength{\baselineskip}{13pt} 
\raggedright				

\end{document}